\documentstyle[aabib,epsfig]{l-aa-ps}
    
    \newcommand{\E}{{\em Einstein}}

    \newcommand{\nli}{\mbox{$ N({\rm Li}) $}}
    \newcommand{\nh}{\mbox{$ N({\rm H}) $}}
    \newcommand{\cq}{\mbox{$ \chi^2$}}

    \newcommand{\feh}{\mbox{[Fe/H]}}

    \newcommand{\dof}{\mbox{degrees of freedom}}

\begin{document}
\thesaurus{6(08.09.2 VY Ari; 08.12.1; 08.01.2; 13.25.5)}

\title{A BeppoSAX observation of the coronal X-ray emission of the active
  binary VY~Ari}

\author{F. Favata\inst{1} \and T. Mineo\inst{2} \and A.~N.
  Parmar\inst{1} \and G. Cusumano\inst{2}}

\institute{Astrophysics Division -- Space Science Department of ESA, ESTEC,
 Postbus 299, NL-2200 AG Noordwijk, The Netherlands
\and
 IFCAI/CNR,
 Via U. La Malfa 153, I-90146 Palermo, Italy }

\offprints{F. Favata (ffavata@astro.estec.esa.nl)
}

\date{Received date; accepted date}

\maketitle 
\begin{abstract}
  
  We present a study of the X-ray coronal emission from the active
  binary VY~Ari, using data from both the Low Energy Concentrator
  Spectrometer (LECS) and the Medium Energy Concentrator Spectrometer
  (MECS) instruments on-board the X-ray satellite BeppoSAX. Using these
  instruments, the X-ray spectrum of VY~Ari can be studied across two
  full decades of energy, from 0.1 to 10\,keV. The spectrum is well
  fit, across the whole spectral range, with an optically thin plasma
  {\sc mekal} model with two discrete temperature components, and with
  an inferred coronal abundance $\feh \simeq -0.4$, corresponding to
  the average typical value of the photospheric abundance in
  RS~CVn-type binaries.


\end{abstract}

\section{Introduction}

Late-type stars in tidally-locked binaries have long since been shown
to be strong coronal X-ray emitters, with much higher X-ray
luminosities than observed in otherwise similar single stars, reaching
values of up to $\simeq 2 \times 10 ^{31}$\,erg\,s$^{-1}$.  This
strong enhancement of the coronal activity has been observed across a
wide range of ages and evolutionary statuses for the components of
tidally locked systems, including Pop.\,I main-sequence and evolved
stars as well as Pop\,II stars. Given the well-demonstrated strong
dependence of coronal X-ray emission on stellar rotation (most likely
through a dynamo mechanism), the enhanced X-ray emission of
tidally-locked binaries with respect to single stars of the same mass
and evolutionary stage can be mostly explained through their high
rotational velocity, induced by the tidal locking mechanism itself.
These systems thus constitute the ideal testbed for studying the
phenomena of stellar activity at rather extreme levels, which are
rarely reached in single stars except during their youngest
evolutionary stages, when the star is still spinning rapidly due to
its residual ``fossil'' angular momentum.

Given their high X-ray luminosities, tidally-locked binaries have been
among the first stellar X-ray sources observed, and among the best
studied to date. The thermal nature of their X-ray emission was shown
using large samples by the \E\ Imaging Proportional Counter (IPC)
observations, and, at the limited spectral resolution of the IPC
detector, most observations are compatible with a simple thermal
structure, modeled through the presence of just one or two plasma
temperature components (e.g.\ \cite{scs+90}). More active stellar
coronae appear, as a trend, to have higher plasma temperatures than
less active ones, and tidally locked binaries obey this rule, so that
their coronal temperatures are generally higher than those of single
stars, with values of up to a few times $10^7$\,K for the hot
component, and a few times $10^6$\,K for the cool component. A similar
picture of coronal emission from tidally-locked binaries emerges from
the observations of the ROSAT All Sky Survey (RASS), with most ROSAT
PSPC spectra also being satisfactorily fit with two isothermal
components (\cite{dlf+93}).

When considered as a class, tidally-locked binaries show some
unexpected peculiarities. Their photospheric metal abundance appears
to be lower than the solar value, with RS~CVn-type systems showing a
typical range of \feh\ comprised between $\simeq -0.8$ and $\simeq
-0.2$, or, equivalently, between $\simeq 0.15$ and $\simeq 0.6$ times
the solar metallicity (\cite{rgp93}; \cite{rgp94}). It is still
unclear what the causes of this widespread photospheric metal
under-abundance are. Tidally locked systems have been variously
classified, with often overlapping classifications (see \cite{fms95c}
for a discussion), depending for example on the rotational period or
on the evolutionary status of the components. Indeed, it is unclear
whether they can be considered together as a class with homogeneous
characteristics, or whether the phenomenological definition of
tidally-locked binary will not group together systems of widely
different characteristics, origins, and evolutionary status.

Thanks to their high X-ray flux, tidally-locked binaries have also
been an obvious target for the higher-spectral resolution CCD
detectors of the ASCA Solid State Spectrometer (SIS) instruments. The
SIS spectra of many coronal systems require a plasma of sub-solar
metallicity to be satisfactorily fit (\cite{whi96}), and this has in
some cases been considered as an indication of the presence of
different metal abundances in the corona with respect to the
photosphere, prompting a debate about the eventual reality of a
so-called ``Metal Abundance Deficiency'' syndrome (or, in brief,
``MAD'' syndrome, \cite{ssd+96}).  However, the case for coronal
under-abundances has often been discussed on the basis of comparison
with the somewhat arbitrary solar abundance values, with the (more
relevant) issue of the relationship between coronal and actual
photospheric abundance having been less thoroughly investigated.
Indeed, many active binaries show lower-than-solar abundance values
both in their photosphere and in their corona, as for example CF~Tuc
(\cite{ssd+96}) or $\lambda$~And (\cite{omp+97}), although for some
single, active stars (i.e. AB~Dor, \cite{mkw+96}) the ASCA- and
EUVE-derived coronal abundance appears to indeed be lower than
photospheric. The first results on coronal emission from the BeppoSAX
LECS detector for both Capella (\cite{fmb+97}) and $\beta$~Cet
(\cite{mfp+97}) show no evidence for sizable coronal under-abundances
with respect to photospheric values.  Many SIS spectra can still be
fit with two discrete temperature components; some of the higher
signal-to-noise spectra do not however yield a satisfactory reduced
\cq\ when fit with such simple models (\cite{whi96}, \cite{dsw+96}),
independently from the assumed coronal abundance. This hints at more
complex temperature structure possibly being present in their corona,
or at problems in the plasma emission codes or in the detector
calibration.

The X-ray satellite BeppoSAX (\cite{bbp+97}) includes four co-aligned
Narrow Field Instruments, of which two sets are of relevance here: the
Low Energy Concentrator Spectrometer (LECS, \cite{pmb+97}), and the
three Medium Energy Concentrator Spectrometers (MECS, \cite{bcc+97}).
The LECS and MECS have imaging capabilities and cover the energy band
0.1--10\,keV and 1.7--10\,keV respectively.

The LECS characteristics make it specially suited for the study of
stellar coronal emission. Its resolution is comparable to the
resolution of CCD detectors at low energies, and it offers access to
the spectral region below $\simeq 0.5$\,keV, where typical coronal
sources emit the largest photon flux. The region below the carbon edge
is specially important as it provides the only essentially line-free
region accessible with a large photon flux in a coronal source. In
addition, the three MECS units provide in the overlapping band an
effective area about three times the one of the LECS alone, with
similar energy and spatial resolution. The larger effective area
obtained by combining the LECS and MECS data at higher energies
allows for example to more effectively study the Fe\,K complex at
$\simeq 6.7$\,keV, a useful diagnostic both of coronal abundance and
of temperature structure in the harder coronal sources. The
unprecedented wide spectral coverage and good spectral resolution, in
particular at the low energies, of the LECS detector, makes it an
effective diagnostic tool for the determination of global coronal
abundance, with its broad band more than offsetting the lower spectral
resolution in comparison with the SIS CCD detectors (\cite{fmp+97}).
The generally lower spectral resolution of the LECS detector with
regard to CCD detectors, however, makes it much less efficient for the
determination of the abundance of individual elements.

Among the sources included in the Science Verification Phase (SVP) of
the BeppoSAX mission, VY~Ari was included as a ``prototype'' coronal
source. VY~Ari (HD~17433) is a bright ($V=6.9$), nearby ($d=21$\,pc),
non-eclipsing SB1 binary system, in which the visible star is
classified as K3--4/V--IV (\cite{bsa+89}), with a so-far unobserved
lower mass companion, which is estimated, on the basis of the mass
function of the system, to be a dM star. The binary orbit is nearly
circular ($e=0.085$), with an orbital period of 13.2\,d. The
photometrically derived rotation period of the primary is $16.4$\,d.
The photospheric lithium abundance (\cite{bsa+89}), at $\nli \simeq
1.0$, is typical for active binaries (\cite{rgp93}), and the surface
gravity of the primary appears consistent with its being slightly
evolved. No determination of its photospheric metal abundance is
available in the literature.

VY~Ari has been observed before in the X-rays by the \E\ IPC detector
and by the ROSAT PSPC detector, both during the ROSAT All Sky Survey
(RASS) and in pointed mode. It has not however so far been observed by
the ASCA satellite. Both the IPC (\cite{scs+90}) and PSPC-RASS
(\cite{dlf+93}) data sets were fit with two-temperature Raymond-Smith
optically-thin plasma models, and yielded best-fit temperatures of
0.22 and 1.55\,keV (IPC) and 0.18 and 1.38\,keV (PSPC), with a ratio
between the emission measures of the soft and hard component of 0.051
and 0.28, respectively. The derived X-ray luminosities were
$1.25\times 10^{30}$ and $2.34\times 10^{30}$ erg\,s$^{-1}$. No
analysis of the ROSAT pointed data (which were taken with the boron
filter in place) has been published thus far.

\section{Observations and data reduction}

The BeppoSAX SVP observation of VY~Ari took place on September 4--6,
1996, resulting in 37\,ks of observing time in the LECS detector and
in 88\,ks of observing time in the MECS detectors, the difference
being due to the LECS being operated only during Earth night at the
time of observation. All the instruments performed nominally during
the observation, and there was no flaring activity, although an
increase in the X-ray count rate during the last BeppoSAX orbit of the
observation suggests the possible on-set of a flare. The number of
photons detected in the last orbit is however too small to allow for
time-resolved spectroscopy to be performed.

The LECS data were reduced through the standard LECS pipeline software
(SAX-LEDAS 1.4.0), which produces a cleaned, linearized photon list,
with energies expressed in pulse-invariant (PI) channels and event
coordinates in RA and Dec. The default screening criteria defined in
the LEDAS system were applied, and inspection of the light curve for
both the source and the local background showed that no additional
screening was necessary.  The MECS data were reduced using the XAS V.\ 
2.0.1 package.  The reduction was performed separately for each unit,
again following the default screening criteria.  The source spectra
were extracted from a circle centered on the source itself of radius
8\,arcmin and 4\,arcmin respectively for the LECS and the MECS.  The
extraction of the LECS spectrum and of the background spectra was
performed using the XSELECT package, and the spectral analysis using
the XSPEC V.\ 9.0 package.  Publicly available matrices (known as
``Dec. 31, 1996'' version) were used for the MECS, while the LECS
response matrix was computed with the LEMAT package (V.\ 3.2.0).  To
subtract the background standard background files obtained by adding
up a set of ``empty sky'' observations were used, extracted from the
same circular region as the source spectra. The source spectra were
re-binned so to have at least 20 counts per re-binned channel, and
channels with energies between 0.1 and 9.0\,keV and between 1.7 and
9.0\,keV were retained for the LECS and MECS detector respectively.
The resulting (background-subtracted) source count rate is
0.28\,cts\,s$^{-1}$, with a background count rate, for the same
region, of 0.05\,cts\,s$^{-1}$ for the LECS, and 0.20\,cts\,s$^{-1}$,
with a background count rate of 0.008\,cts\,s$^{-1}$ for the set of
three MECS detectors taken together.

\section{Results}

The resulting spectra from the LECS and from the MECS detectors were
simultaneously fit with an optically-thin plasma emission model with
two discrete temperature components and freely varying global metal
abundance. The {\sc mekal} plasma emission model (\cite{mkl95}) was
used throughout, as implemented in XSPEC 9.0. The effect of
interstellar absorption was included, by adding a {\sc wabs}
multiplicative component to the spectral model (implementing the
\cite{mm83} interstellar absorption model). Given the current
uncertainties on the absolute calibration of the various detectors,
the relative normalization of each MECS detector with respect to the
LECS has been left as an additional free parameter in the fit. The fit
converges to a relative normalization of the MECS detectors $\simeq
30$\% higher than the LECS, in line with the expected calibration
uncertainties at this stage. Due to the longer exposure time, as well
as to the larger effective area, the MECS spectra allow a much better
constraint to be placed on the flux from Fe\,K complex.  Detection of
the Fe\,K complex supplies a diagnostic of the metallicity of the
emitting plasma which is independent from the main diagnostic
available to LECS spectra, i.e. the balance between the essentially
line-free low-energy continuum (below the carbon edge) and the
line-rich region around 1\,keV, mostly due to the Fe\,L line complex.

\begin{figure}[htbp]
  \begin{center}
    \leavevmode
    \epsfig{file=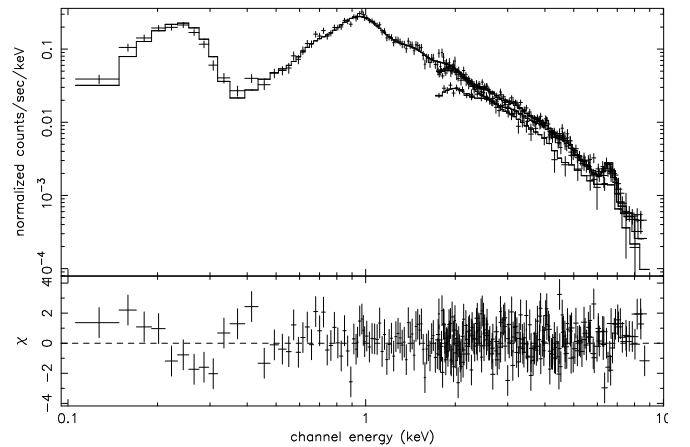, width=9.cm, bbllx=5pt, bblly=30pt,
    bburx=687pt, bbury=475pt, clip=}
  \end{center}
  \caption{
    The observed BeppoSAX LECS and MECS (for $E \ge 1.7$\,keV)
    spectrum of VY~Ari, together with the best-fit two-temperature
    {\sc mekal} model spectrum. To reduce visual clutter the data have
    been re-binned, in the plot, by an additional factor of two with
    respect to the binning scheme used for the fit}
  \label{fig:highz}
\end{figure}

The resulting best-fit model is shown, together with the observed
spectrum and the fit residuals, in Fig.~\ref{fig:highz}.  The fit has
a reduced \cq\ of 1.09 (with 523 \dof). The best-fit temperatures are
$0.85\pm 0.04$ and $2.1 \pm 0.08$\,keV, with a ratio between the two
emission measures (soft/hard) of 0.44. The best-fit metal abundance is
$0.37\pm 0.04$ times the solar value (corresponding to $\feh \simeq
-0.4$), and the best-fit implied hydrogen column density is $(4.3 \pm
0.9) \times 10^{19}$\,cm$^{-2}$.

Published values for the column density toward VY~Ari are somewhat
inconsistent. \cite*{fhj+91} report on column density measurements
toward the star $\rho$~Ari, which is close to VY~Ari (at a projected
distance of 14\,deg in the sky) and at a similar distance (25\,pc)
Based on the column density of Na\,{\sc i}, these data yield an upper
limit to the value of \nh\ of $1.2\times 10^{19}$\,cm$^{-2}$. Another
upper limit to the column density of $2.0\times 10^{19}$\,cm$^{-2}$
has been derived by \cite*{djp95} for a distance of 20\,pc in the sky
direction of VY~Ari, based on the observation of ROSAT EUV Wide Field
Camera sources.  A value of $2.2\times 10^{19}$\,cm$^{-2}$ is obtained
by \cite*{dlf+93} from a spectral fit to the RASS data. \cite*{dbh96}
quote a much lower value of $1.0\times 10^{18}$\,cm$^{-2}$, derived
from an analysis of the EUVE spectrum.

The LECS-derived best-fit value for the column density appears to be
on the high side with respect to previously derived values, which are
however not fully consistent with each other. Given the nature of the
LECS spectral response, source metallicity and interstellar absorption
are not wholly uncorrelated, with a low metallicity spectrum
potentially mimicking the effects of high interstellar absorption. To
assess the possible influence of such an effect we repeated the fit
fixing the interstellar absorption at $1.0 \times 10^{19}$\,cm$^{-2}$,
a value significantly lower than the best-fit one, but more consistent
with the previously available estimates. The fit in this case
converges to a slightly worse reduced \cq\ value of 1.13, with a
best-fit metallicity of $0.51\pm0.04$ times solar, higher, as
expected, than the best-fit metallicity deduced from a fit converging
to a higher interstellar absorption column density. The best-fit
temperatures do not appreciably change, being 0.81 and 2.0\,keV.

There are unfortunately no photospheric abundance measurements
available for VY~Ari, and the best-fit coronal metallicity cannot thus
be directly compared with the photospheric value. A comparison can
however be made with typical photospheric metallicity values for
tidally-locked binaries. \cite*{rgp93} and \cite*{rgp94} have derived
photospheric lithium and iron abundance values for a large set of
active binaries, through analysis of a small segment of their
high-resolution optical spectrum, showing that the photospheric iron
abundance of RS~CVn-type systems, as a class, is significantly lower
than the solar one. In particular, the typical range of \feh\ values
observed in RS~CVn binaries lies between $\simeq -0.8$ and $\simeq
-0.2$. Thus, the best-fit coronal metallicity for VY~Ari, which
corresponds to $\feh \simeq -0.4$, is close to the average of the
photospheric abundance values observed for this class of objects.

The best-fit temperatures for the BeppoSAX spectra are rather high
when compared with the IPC and PSPC best-fit temperatures. However,
the temperatures are comparable with typical coronal temperatures of
active binaries derived from the analysis of ASCA SIS spectra.
Similar differences between the coronal temperatures derived from the
BeppoSAX spectra and from previous analyses have also been seen in the
case of Capella (\cite{fmb+97}), and are, at least in part, likely to
be due to differences in the plasma emission codes used in the
analysis, although the effect of the much more limited energy range
observed by either the \E\ IPC or the ROSAT PSPC is also likely to
play a role, somewhat ``forcing'' the analysis toward lower effective
temperatures for the hotter plasma components. For a discussion of the
effects of the detector characteristics on the derived
``temperatures'' see for example \cite*{msg+86}, \cite*{wbl95} and
\cite*{dlf+93}.  The X-ray luminosity in the 0.16--3.5\,keV band
resulting from the analysis of the BeppoSAX spectra is $1.1\times
10^{30}$ erg\,s$^{-1}$, i.e.  slightly lower than the IPC- and
PSPC-derived luminosities.  However, given the typical range of
intrinsic source variability observed in this class of objects, the
agreement between the measured source luminosities appears to be good.

\section{Discussion}

The first broad-band coronal spectrum obtained with the LECS and MECS
detectors on-board the BeppoSAX satellite is well fit with a
two-temperature, optically-thin plasma model, using the {\sc mekal}
plasma emission code. With a spectrum of moderate signal-to-noise
ratio and spectral resolution, such as the one discussed here, with
about 10\,kcts in the LECS spectrum, and about 17\,kcts in the three
MECS spectra, a simple two-temperature model gives a satisfactory
phenomenological description of the X-ray spectrum across two decades
of energy --- between 0.1 and 10\,keV. The resulting coronal
abundance, at $\feh\simeq -0.4$ is typical for the photospheric
abundance of RS~CVn-type binaries.

In this work we assume a two temperature model is a good
representation of the coronal structure of VY~Ari. As shown by
\cite*{fbd+97}, simple \cq\ minimization approaches, as currently
implemented in X-ray spectral fitting packages, may not always
converge, when applied to moderate resolution spectra, toward more
complex temperature structures, even when such structures constitute a
better (i.e. lower \cq) representation of the observed spectrum. The
differential emission measure (DEM) derived by EUVE has been shown in
these cases to be a good starting point for finding a more realistic
solution. We will, in a future paper, analyze the VY~Ari spectrum in a
similar fashion to that performed for Capella by \cite*{fmb+97}.

\acknowledgements{We have used the Simbad database in the course of
  the present work. We would like to thank A.~Maggio and R.~
  Pallavicini for the many useful comments and suggestions.  IFCAI is
  supported by the Italian Space Agency (ASI) in the framework of the
  BeppoSAX mission. The BeppoSAX satellite is a joint Italian and
  Dutch program.}

\end{document}